# Honeycomb lattice iridate on the verge of Mott-collapse

Yuya Haraguchi[*], and Hiroko Aruga Katori

Department of Applied Physics and Chemical Engineering, Tokyo University of Agriculture and Technology, Koganei, Tokyo 184-8588, Japan

E-mail: chiyuya3@go.tuat.ac.jp

**Abstract**

A new honeycomb lattice iridate $(La,Na)IrO_3$ ($\approx LaNaIr_2O_6$) is successfully synthesized from the spin-orbit coupled Mott insulator $Na_2IrO_3$ by replacing the interlayer $Na^+$ ions with $La^{3+}$ ions. $(La,Na)IrO_3$ shows a finite Sommerfeld term in heat capacity and a $-\ln T$ dependence of resistivity, indicating a realization of a metallic state driven by a Mott collapse. Furthermore, crystal structure analysis reveals the formation of Ir zig-zag chains with metal-metal bonding, increasing kinetic energy resulting in the Mott collapse. This observation would be due to a Mott collapse induced in a $J_{eff} = 1/2$ spin-orbit coupling Mott insulator with an Ir honeycomb lattice by topochemical control of the ionic configuration.

## I. Introduction

Materials consisting of heavy $5d$ transition-metal ions have a great potential for the emergence of unprecedented electronic states owing to the strong spin-orbit coupling (SOC), comparable to the electron kinetic energy Coulomb repulsion [1]. Especially in a low spin $d^5$ ion like $Ir^{4+}$, strong SOC splits the $t_{2g}$ band of $5d$ electrons into half-filled $J_{eff} = 1/2$ and filled $J_{eff} = 3/2$ bands. Moreover, in the half-filled $J_{eff} = 1/2$ bands, the correlation effect is prominent even for relatively weak Coulomb repulsion, which results in the realization of novel electronic solids called "SOC Mott insulators" [1,2].

Such SOC Mott insulators provide us with an opportunity to find exotic magnetic ground states driven by a Kitaev-type bond-sensitive anisotropic interaction, which is emergent in the spin-orbit entanglement of $J_{eff} = 1/2$ wave function [3]. When the Kitaev-type interaction is arranged on a honeycomb lattice, the ground state is exactly solved as a quantum spin liquid [4]. In realistic materials, the non-Kitaev term destabilizes the spin liquid state, which results in the magnetic ordering as a ground state [5-14]. It has been reported in the previous studies that $Na_2IrO_3$ and $RuCl_3$, in which $J_{eff} = 1/2$ pseudospins form a honeycomb lattice, show the magnetic ordering with a zig-zag structure originated from a competition between Kitaev and Heisenberg interactions [8, 12]. Especially in $RuCl_3$, a half-integer quantized anomalous thermal Hall effect has been observed when the magnetic order is suppressed under magnetic fields, which strongly supports a realization of Kitaev spin liquid [15]. In addition, it is reported that $H_3LiIr_2O_6$, which is prepared by a soft chemistry technique replacing interlayer lithium ions of $Li_2IrO_3$ with protons, shows a spin-liquid behavior as a ground state [16].

The Mott insulating phase is characterized by a metallic phase with a strong electron correlation in its vicinity. There are two methods to induce the Mott transition: bandwidth-controlling and filling-controlling. The former increases the kinetic energy by compressing the lattice, and the latter reduces the effective on-site Coulomb repulsion by chemical substitution. It has been argued that the quantum spin liquid state may be closely related to the expression mechanism of high-temperature superconductors [17]. Therefore, the Mott transition in the Kitaev magnet is expected to lead to an exotic superconducting phase in the vicinity of the Kitaev spin liquid [18, 19]. However, no realization of Mott transition in Kitaev magnetic material has been reported so far in the previous studies. The pressure-applied Kitaev magnets (α-$Li_2IrO_3$, $Na_2IrO_3$, and $RuCl_3$) do not show metallization but dimerization [20-23]. Moreover, there have been attempts to control the Kitaev magnet by inserting alkali metals between the layers. However, all of them have been reported to cause charge disproportionation instead of the Mott transition [24, 25].

Here, we report on the synthesis of a new honeycomb lattice iridate $(La,Na)IrO_3$ via a topochemical ion-exchange reaction and the investigation of its physical properties. The electronic state of $(La,Na)IrO_3$ is beyond a $J_{eff} =1/2$ Mott insulator, evidenced by the Sommerfeld heat capacity at low temperatures, weak Curie-Weiss-like magnetic susceptibility, and almost temperature-independent electric resistivity with $-\ln T$ dependence. This behavior would be driven by a Mott collapse of a Kitaev magnet induced by a topochemical control of the ionic configuration.

## II. Experimental Procedure

The precursor $Na_2IrO_3$ was obtained by the conventional solid-state-reaction method according to the previous report [7]. This precursor was ground well with three-fold excess of $La(NO_3)_3$ in an Ar-filled glovebox, sealed in an evacuated Pyrex tube, and reacted at 300°C for 50 hours following the ideal reaction: $2Na_2IrO_3 + La(NO_3)_3 \rightarrow LaNaIr_2O_6 + 3NaNO_3$. The unreacted starting material $La(NO_3)_3$ and the by-product $NaNO_3$ were removed by washing the sample with distilled water. The product was characterized by powder X-ray diffraction (XRD) experiments in a diffractometer with Cu-Kα radiation. The cell parameters and crystal structure were refined by the Rietveld method using the RIETAN-FP v2.16 software [26]. The temperature dependence of the magnetization was measured using a magnetic property measurement system (MPMS; Quantum Design) equipped at the Institute for Solid State Physics at the University of Tokyo.

**Table 1** Crystallographic parameters for $(La,Na)IrO_3$ (space group: *Cmcm*) determined from powder x-ray diffraction experiments. For structural analysis, the composition is assumed to be Na:La:Ir = 1:1:2 based on the ideal interlayer $3Na^+ \leftrightarrow La^{3+}$ exchange reaction. The obtained orthorhombic lattice parameters are $a = 9.1568(2)$ Å, $b = 5.2894(2)$ Å, $c = 9.2813(4)$ Å. *B* is the atomic displacement parameter, and *g* is the occupancy factor.

| atom | g | x | y | z | B(Å) |
|---|---|---|---|---|---|
| Na | 1 | 0 | 0 | 0 | 1.8(2) |
| La1 | 0.215(1) | 1/6 | 5/6 | 1/4 | 2.2(1) |
| La2 | 0.119(1) | 1/3 | 0 | 1/4 | 2.2(1) |
| La3 | 0.331(1) | 0 | 0 | 1/4 | 2.2(1) |
| Ir | 1 | 0.3229(2) | 0 | 0 | 0.16(1) |
| O1 | 1 | 0.1726(7) | 0.227(1) | 0.122(1) | 0.4(1) |
| O2 | 1 | 0 | 0.667(2) | 0.116(1) | 0.4(1) |

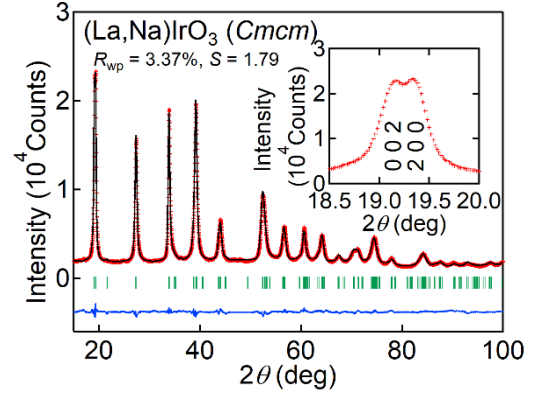

**Figure 1** Powder x-ray diffraction patterns of $(La,Na)IrO_3$. The observed intensities (red), calculated intensities (black), and their differences (blue) are shown. Vertical bars indicate the positions of Bragg reflections. The inset shows the enlarged view near the lowest angle peaks 002 and 200.

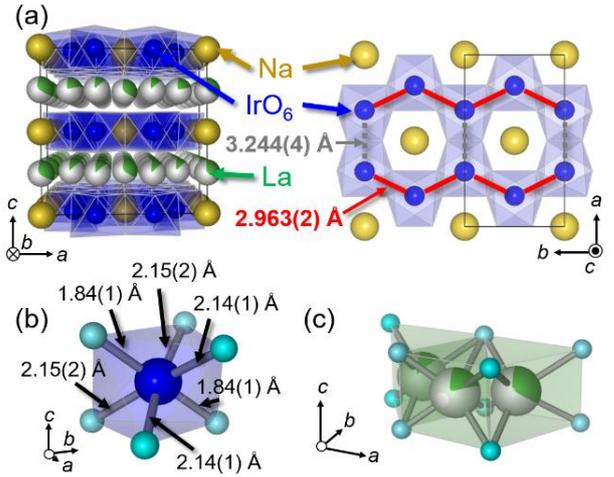

**Figure 2** (a) Crystal structure of $(La,Na)IrO_3$ viewed along with the layers (left) and perpendicular to the layer (right) with Ir networks on the honeycomb lattice, where two Ir-Ir bonds with different lengths are shown differently. (b) The local environment of $IrO_6$ octahedron with the three different Ir-O bond lengths. (c) The local environment of $LaO_6$ trigonal prism. The VESTA program is used for visualization [27].

The temperature dependence of the electric resistivity was measured using a conventional dc four-probe method with a hand-made apparatus. The temperature dependence of the specific heat was measured using a conventional relaxation method with a physical property measurement system (PPMS; Quantum Design) equipped at the Institute for Solid State Physics at the University of Tokyo.

## III. Results

The powder XRD pattern of (La,Na)IrO$_3$ is shown in Fig. 1. All the peaks can be indexed with the space group of *Cmcm* with the orthorhombic lattice constants $a$ = 9.1568(2) Å, $b$ = 5.2894(2) Å, $c$ = 9.2813(4) Å. As shown in the inset of Fig. 1, most XRD peaks have merged because the $a$ and $c$ parameters take relative values, but separation can be seen when magnified. An interlayer 3Na$^+$↔La$^{3+}$ exchange reaction has been reported in the reaction in which La$_{0.3}$CoO$_2$ is generated from Na$_{0.9}$CoO$_2$ using La(NO$_3$)$_3$ [28]. Thus, we constructed an initial structural model after an interlayer 3Na$^+$↔La$^{3+}$ exchange reaction similar to La$_{0.3}$CoO$_2$, assuming that the Na ions occupying the voids between the IrO$_6$ octahedra are not exchanged by the La ions, and refined the atomic positions by the Rietveld method. The powder pattern is perfectly reproduced using the crystal structure displayed in Fig. 2(a). The details of the refinement parameters are given in Table 1. There is only one crystallographic site for the Ir atom, which is octahedrally coordinated by six oxygen atoms. The IrO$_6$ octahedra are linked by their common edges, resulting in a formation of a two-dimensional layer. Since 1/3 of the Ir atoms are regularly replaced by Na atoms from a triangular lattice, a Na$_{1/3}$Ir$_{2/3}$O$_2$ = NaIr$_2$O$_6$ layer containing a honeycomb lattice made of Ir$^{4+}$ ions is generated. The NaIr$_2$O$_6$ layers are separated from each other by a block layer consisting of La$^{3+}$ ions. This structure supports that only interlayer Na$^+$ ions sandwiched between the NaIr$_2$O$_6$ layers in the precursor Na$_2$IrO$_3$ = Na$_3$NaIr$_2$O$_6$ are selectively exchanged for La$^{3+}$ ions. This interlayer ion exchange is similar to the synthesis process of delafossite-type honeycomb lattice iridates (H/Cu/Ag)$_3$LiIr$_2$O$_6$ [16,29,30].

Attempts at producing intermediate stoichiometries (i.e., Na$_{2-3x}$La$_x$IrO$_3$, $x \neq 0.5$) by reducing the amount of La(NO$_3$)$_3$ to be added were unsuccessful and instead produced two-phase samples containing a mixture of Na$_2$IrO$_3$ and (La,Na)IrO$_3$. Further, no phase width was observed in (La,Na)IrO$_3$, as reactions conducted with more significant excess La(NO$_3$)$_3$ produced (La,Na)IrO$_3$ with no observable difference in lattice parameters and unknown decomposed impurities. These facts indicate no irregularity in the chemical composition of (La,Na)IrO$_3$. Unfortunately, inductively coupled plasma measurement for clarifying a detailed chemical composition cannot be attempted because (La,Na)IrO$_3$ is not soluble in strong acids. Generally, energy dispersive X-ray spectroscopy (EDX) measurement using a powder sample is less accurate because the analysis method assumes a flat surface and homogeneity of a few micrometers cubic. Indeed, EDX measurement yields a relatively large error. In addition, there are some measurement areas where no cations other than La are detected, which indicates that there are impurities on the sample surface undetectable by XRD. Therefore, the establishment of single crystal growth and detailed chemical composition analysis are future issues. However, the bond valence sum calculation for Ir ions from the refined structural parameters yields +4.117, consistent with the expected valence of +4, indicating that the chemical composition and valence is nearly ideal. Hence, we conclude that this topochemical reaction does not involve electron/hole doping. Therefore, the data analysis of the physical properties in the following sections is based on the chemical composition of La$_{0.5}$Na$_{0.5}$IrO$_3$ = La[NaIr$_2$O$_6$], assumed redox-neutral ion exchange reactions and no exchange of Na ions located in the center of the hexagon of the honeycomb lattice.

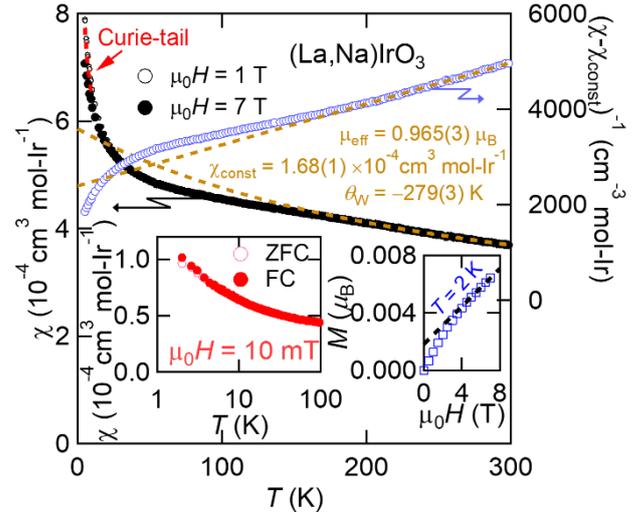

**Figure 3** Temperature dependence of magnetic susceptibility $\chi$ and inverse susceptibility corrected by the temperature-independent term $\chi_{const}$ as $(\chi - \chi_{const})^{-1}$ of a powder sample of (La,Na)IrO$_3$ at 1 and 7 T. The left inset shows $\chi$ data measured under 10 mT. Measurements were conducted upon heating after zero-field cooling (ZFC) and field cooling (FC) for 10 mT and only FC conditions for the others. The right inset shows the isothermal magnetization at 2 K.

The absence of the threefold axis in the orthorhombic unit cell distorts the Ir honeycomb network, where two types of Ir-Ir bonds with different lengths are shown in the right part of Fig 2(a). One of the bonds making a zig-zag chain extending toward the b-axis direction has shorter Ir-Ir distances of 2.963(2) Å, and the other comprising the zig-zag chain has longer Ir-Ir distances of 3.244(4) Å. Each Ir ion is surrounded by six oxygens with 2-short-4-long-type coordination, as shown in Fig. 2(b), indicating that Ir ions are subjected to a tetragonal-like crystal field rather than a trigonal-like one expected in the honeycomb lattice compounds. Li$^+$/Na$^+$ ion in (Li/Na)$_2$IrO$_3$ has the octahedral coordination and H$^+$/Cu$^+$/Ag$^+$ ions in (H/Cu/Ag)$_3$LiIr$_2$O$_6$ has the linear coordination around

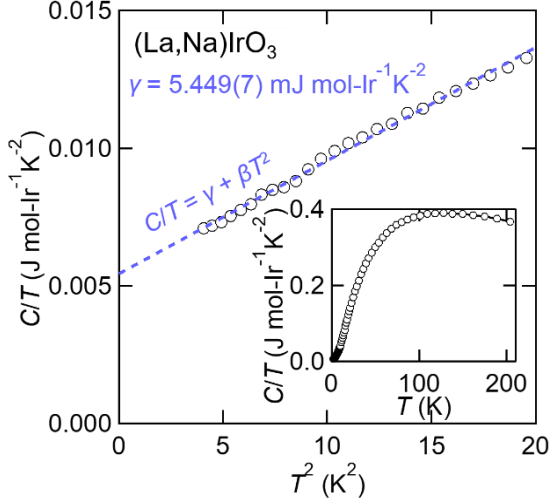

**Figure 4** The heat capacity divided by temperature $C/T$ as a function of $T^2$ below 20 K$^2$ with a fit to the equation $C/T = \gamma + \beta T^2$ (blue dashed line). The inset shows $C/T$ as a function of $T$ up to 200 K, indicating no phase transition.

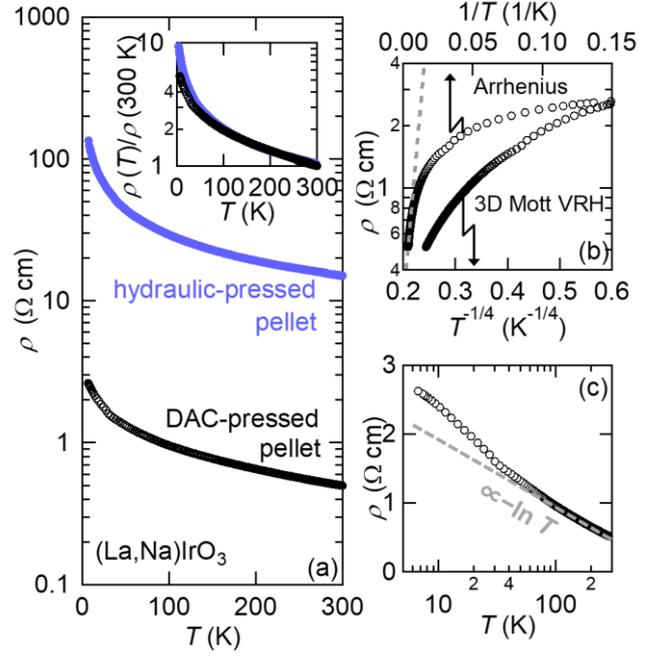

**Figure 5** (a) Temperature dependence of electric resistivity $\rho$ of a cold-pressed pellet of $(La,Na)IrO_3$ prepared by applying a pressure of 2 GPa or 25 MPa using a diamond anvil cell (DAC) and a hydraulic press machine, respectively. The inset shows the temperature dependence of normalized resistivity $\rho(T)/\rho(300\ K)$. (b) The black circles show the logarithmic $\rho$ vs. $1/T$ plot (Arrhenius thermal activation model), where the broken blue line indicates a fit at high temperatures (> 200 K) to thermally activated Arrhenius behavior yielding a gap of $\Delta = 26.9(2)$ meV. The blue circles show the logarithmic $\rho$ vs. $T^{-1/4}$ plot (Mott 3D variable range hopping model). (c) The $\rho$ vs. $\ln T$ plot, exhibiting good linearity over a wide temperature range above 50 K. The gray dashed line represents a fit to $\rho(T) \sim -\ln T$.

[6,16,29,30]. On the other hand, $La^{3+}$ ions occupy trigonal prismatic coordination with oxygen ions, as shown in Fig. 2(c).

Figure 3(a) shows magnetic susceptibility $\chi$ of the powder sample of $(La,Na)IrO_3$ measured at magnetic fields of 1 and 7 T. Above 200 K, the $\chi$ data is fitted well with using the extended Curie-Weiss function,

$$\chi = (N\mu_B^2/3k_B)\mu_{eff}^2/(T-\theta_W) + \chi_{const}, \quad (1)$$

where $\mu_{eff}$ is the effective magnetic moment, $\theta_W$ is the Weiss temperature, and $\chi_{const}$ is the temperature-independent term. This fitting yields the parameters $\mu_{eff} = 0.965(3)\ \mu_B$, $\theta_W$ −279(3) K, and $\chi_{const} = 1.68(1) \times 10^{-4}\ cm^3\ mol\text{-}Ir^{-1}$. The value of $\mu_{eff}$ is significantly lower than the value of 1.73 $\mu_B$/Ir for $J_{eff} = 1/2$. This reduction may indicate an itinerancy of electrons or a strong hybridization between Ir ion and oxygen. The relatively large $\chi_{const}$ may be the stoner enhancement in the Pauli paramagnet. Note that there is an upturn in low temperatures, ascribed to a contribution of impurity spins of 0.4% $Ir^{4+}$ local moment. As shown in the inset of Fig. 3, a trace of spin glass-like contribution is observed under a low magnetic field of $\mu_0 H = 10$ mT. The difference between zero-field cooled (ZFC) and field cooled (FC) magnetization is less than 4% and sample dependent, strongly suggesting that the glass-like behavior is due to minority spins or impurities.

The inset of Fig. 4 shows the heat capacity of $(La,Na)IrO_3$, where no phase transition is observed in the measured temperature range, which agrees with $\chi$. The main panel of Fig. 4 shows low-temperature $C/T$ data as a function of $T^2$, of which linear dependence with a finite intercept indicates Fermi liquid behavior. The $C/T$ data were fit to the equation $C/T = \gamma + \beta T^2$, where $\gamma$ and $\beta$ model the electronic and phonon contributions to $C$, respectively. The finite $\gamma$ value of 5.449(7) mJ mol-Ir$^{-1}$ K$^{-2}$ indicates a finite density of states (DOS) at the Fermi level. It is comparable to other semimetallic iridates, for example, hyperkagome $A_3Ir_3O_8$ ($A$ = Li, Na) ($\gamma$ = ~5.0 mol-Ir$^{-1}$K$^2$) [31,32] and perovskite $SrIrO_3$ ($\gamma$ = 4.4 mol-Ir$^{-1}$K$^2$) [33]. Thus, this observation evidences that $(La,Na)IrO_3$ is metallic/semimetallic. In the metallic/semimetallic scenario, the observed weak Curie-Weiss-like behavior of $\chi$ in $(La,Na)IrO_3$ would be explained by the theories of itinerant electron systems predicting the temperature dependence of the amplitude of local spin density in itinerant magnets also results in Curie-Weiss behavior [34-36]. The Debye temperature $\Theta_D$ is estimated to be 381 K from $\beta = 0.351(1)$ mJ

mol-Ir$^{-1}$K$^{-4}$, a typical value of oxide materials with metallic conductivity.

Figure 5(a) shows the temperature dependence of electric resistivity ρ of two cold-pressed pellets of (La,Na)IrO$_3$, which were prepared by applying a pressure of 2 GPa or 25 MPa using a diamond anvil cell (DAC) and a hydraulic press machine, respectively. The effect of grain boundaries is seemingly reduced by applying higher pressure. However, both ρ data show almost identical temperature changes except for their absolute values. In addition, as shown in the inset of Fig. 5(a), the temperature dependence of normalized resistivity ρ(*T*)/ρ(300 K) is roughly consistent over the entire temperature range. Thus, the observed temperature dependence is intrinsic. In the following, we will discuss the ρ data of the DAC-pressed pellet. It increases with decreasing temperature suggesting semiconducting behavior. As shown in Fig. 5(b), a forced fitting between 200 to 300 K to the Arrhenius model ln ρ ~ $T^{-1}$ yields a minimal gap of *Δ* = 26.9(2) meV. However, the data for most of the temperature range could not be reproduced at all. To investigate the possibility of variable range hopping (VRH) conduction [37], we replotted the following equation: lnρ vs. $T^{-1/(d+1)}$, where *d* is the dimensionality *d* = 3 found no linearity but convex upward behavior in all temperature ranges. At lower dimensions, i.e., *d* = 2 and 1, the lnρ data have a more upward convex curvature and less linearity. Therefore, the VRH model, as expected in Mott insulators, cannot explain the temperature dependence of ρ in (La,Na)IrO$_3$. On the other hand, as shown in Fig. 5(c), the ρ data follows the −ln*T* dependence asymptotically over a wide temperature range of 50< *T*/K < 300. This temperature dependence is reminiscent of the weak localization effect due to disorder expected in two-dimensional electron gases, thin films [38], and amorphous alloys [39]. However, this explanation is not appropriate because this system is not a dilute electron or thin film. Therefore, a detailed discussion will be given in the next chapter.

## IV. Discussion

(La,Na)IrO$_3$ shows signatures of metallic ground state have been observed in the *T*-linear heat capacity and the Pauli paramagnetism. The temperature dependence of the resistivity cannot be explained in the categories of ordinary semiconductors or Mott insulators. The −ln*T* dependence reminds us first of the disorder-induced two-dimensional weak localization effect observed in thin film materials [40] or the Kondo effect in *f*-electron systems [41]. In the present case, this system is not an *f*-electron system and is expected to be an effect of weak localization in a two-dimensional system. The elastic scatterings of electrons by disorders or imperfections occur several times within a metallic domain in which phase coherence of the wavefunction is preserved. Hence, the observed −ln*T* dependence in ρ suggests a realization of the metallic state.

Weak localization by disorder is somewhat reminiscent of systems on the boundaries of Anderson localization [42]. The disorder of La sites revealed by structural analysis may be sufficient to cause Anderson localization. However, the 5-fold increase in ρ from 300 to 4 K is too slight compared to Si$_{1-x}$P$_x$ [43] and Ce$_{2.67}$S$_4$ [44], known as Anderson localization examples. In addition, the Sommerfeld term in heat capacity of (La,Na)IrO$_3$ is nearly 50 times larger than that in Si$_{1-x}$P$_x$ [45]. Therefore, the Anderson localization scenario seems not to be a match quantitatively. Therefore, it is necessary to consider mechanisms other than Anderson localization. So far, two-dimensional weak localization has only been observed in thin metal samples and amorphous alloys. Therefore, it is not yet clear whether it is appropriate to apply to bulk materials. Further studies involving a high-quality single crystal are required to fully characterize the ρ(*T*) behavior of (La,Na)IrO$_3$.

As an alternative model to explain the weak localization *without* the disorder, it may be helpful to consider the "nonmetallic" metal states. Because of the strong spin fluctuations due to the proximity to the metal-insulator transition, even metals are predicted to exhibit non-metal resistivities [46]. The spin fluctuation forms an energy gap, and the Fermi surface is buried in the gap, resulting in weak localization behavior. This mechanism is similar to the Kondo insulator. Since it is widely known that the Kondo effect generates a −ln*T* dependence of electrical resistance, spin fluctuations near metal-insulator transition in the nonmetallic metal can also generate a similar temperature dependence of electrical resistance. In other words, the electrons that begin to localize near the metal-insulator transition interact with the conduction electrons and behave like the Kondo effect. This speculation will be clarified theoretically in the future. However, even in this situation, the *T*-linear heat capacity should appear as a metallic character. The nonmetallic metal state was first observed in FeCrAs [47, 48]. In addition, such a nonmetallic metal behavior has recently been observed in some oxides, for example, Lu$_2$Rh$_2$O$_7$ [49] and Cu$_3$LiRu$_2$O$_6$ [50]. Therefore, a similar weak localization state may be realized in (La,Na)IrO$_3$ near the Mott transition.

Here, we estimate the electron correlations in (La,Na)IrO$_3$. The Wilson ratio $R_W = (\pi^2 k_B^2/3\mu_B^2)(\chi_0/\gamma) \sim 72.95(\chi_0/\gamma)$ is often used to estimate the magnitude of electron correlation in itinerant electron systems [51]; $R_w = 1$ in the case of the Fermi gas without correlation. With an enhancement of electron correlation, $R_W$ increases from 1. The $\chi_0$ value is estimated to be approximately 6×10$^{-4}$ cm$^3$ mol-Ir$^{-1}$ using an ordinate intercept of the Curie-Weiss fitting curve, resulting in a value of $R_w \sim 8$. This value remains almost the same when using the $\chi_0$-value of 6.59(3)×10$^{-4}$ cm$^3$ mol-Ir$^{-1}$ obtained by linear fitting the high magnetic field region of the isothermal magnetization process [see the right inset of Fig. 3]. Such a significant $R_W$ value is also observed in itinerant magnets such as element Pd ($R_w$ ~ 7) [52] and $Ca_{0.5}Sr_{1.5}RuO_4$ ($R_w$ ~ 40) [53]. Hence, the observed significant $R_w$ value possibly indicates that (La,Na)IrO$_3$ is a stoner-enhanced, strongly correlated electron system similar to these examples. Note that the $R_W$ value could be overestimated due to a van Vleck contribution. The $\chi_{const}$ = $1.68\times10^{-4}$ cm$^3$ mol-Ir$^{-1}$ of (La,Na)IrO$_3$, however, is larger than the $\chi_{const}$ ~ $2.66\times10^{-5}$ for the precursor material Na$_2$IrO$_3$. Thus, it is reasonable to think that the observed $R_w$-value cannot be explained only by van Vleck paramagnetism. Nevertheless, the interlayer 3Na$^+$ ↔ La$^{3+}$ ion-replacing metallizes the SOC Mott insulator Na$_2$IrO$_3$ powerfully indicates that (La,Na)IrO$_3$ is a strongly correlated electron system on the verge of the Mott transition.

Finally, we discuss the origin of Mott collapse in (La,Na)IrO$_3$. All the Ir$^{4+}$ honeycomb lattice compounds reported so far are Mott insulators. Under high pressure, it has been found that Li$_2$IrO$_3$ and Na$_2$IrO$_3$ show a dimerization between short Ir-Ir bonds, which has been reasonably explained by the increase of orbital hybridization to form molecular orbitals between short Ir-Ir bonds as the lattice constant is reduced [20-23]. Also, in (La,Na)IrO$_3$, two types of Ir-Ir bonds with long and short distances exist as described above. Thus, between the short Ir-Ir bonds, orbital hybridization should be formed. However, these short Ir-Ir bonds form an infinite zig-zag chain rather than orphan Ir-Ir dimers, observed in Li$_2$IrO$_3$ and Na$_2$IrO$_3$. Similar zig-zag chain formation is observed in pressurized Li$_2$RhO$_3$, which bears an electronic structure of d$^5$ similar to the iridates [54]. Therefore, it is reasonable to assume that a zig-zag chain formation increases the kinetic energy of the electrons, then Mott collapse.

As another possibility, a low-symmetric crystal field realizes a Mott collapse. In order to clarify how the pseudo-tetragonal crystal field found in the crystal structure analysis affects the electronic structure of SOC Mott insulators with a honeycomb geometry, further development of the theoretical investigation is needed. Photoconductivity and photoelectron spectroscopy measurements conclusively confirm whether (La,Na)IrO$_3$ is metallic or not. However, obtaining essential data from these measurements is difficult as only powder samples have been obtained. Therefore, establishing a single crystal synthesis method of (La,Na)IrO$_3$ to clarify the ground state is a future issue.

## V. Summary

We successfully synthesized a novel honeycomb lattice iridate (La,Na)IrO$_3$ via a topochemical manipulation. The measurements of physical properties suggest that (La,Na)IrO$_3$ is no longer a spin-orbit coupled Mott insulator. Signatures of a Mott collapse have been observed in a finite Sommerfeld coefficient of the electronic heat capacity, a small effective magnetic moment, and $-\ln T$ dependence of resistivity.

Considering the relationship between the crystal structure and the physical properties, the origin of a Mott collapse would be a zig-zag chain formation with short Ir–Ir bonds. The observed Mott collapse by topochemical cation substitution is expected to bridge the physics of SOC Mott insulators and strongly correlated electron systems.

## Acknowledgments

We thank A. Yamamoto for the sample treatment of the resistivity measurement using DAC at the Shared Equipment Centers of Shibaura Institute of Technology. This work was supported by the Japan Society for the Promotion of Science (JSPS) KAKENHI Grants Nos. JP22K14002, JP19K14646, and JP21K03441. Part of this work was carried out by the joint research in the Institute for Solid State Physics, the University of Tokyo.